\documentclass[11pt,fleqn]{article}
\usepackage{psfig}
\title{Transport Control of Eyring-Fluids along a Transversely-Corrugated Nanoannulus}  %
\author{Zhu Guang Hua} 
\date{Department of Physics, Wulumuqi Toudiban (Transfer Centre), Xinjiang 830000, China}
\begin{document}           
\maketitle
\begin{abstract}
The volume flow rates of Eyring-fluids inside the wavy-rough
nanoannulus were obtained  analytically (up to the second order)
by using the verified  model and boundary perturbation method. Our
results show that the wavy-roughness could enhance the flow rate
especially for smaller forcing due to the larger surface-to-volume
ratio and slip-velocity effect. Meanwhile, the phase shift between
the outer and inner walls of nanoannuli could tune the transport
of Eyring-fluids either forward or backward when the
wavy-roughness of a nanoannulus is larger enough. Our results
could be applied to the flow control in nanofluidics as well as
biofluidics.  \newline

\end{abstract}
\doublerulesep=6mm        
\baselineskip=6mm
\oddsidemargin-1mm         
\bibliographystyle{plain}
\section{Introduction}
Solutions of linear chain structures exhibit interesting
rheological properties such as the  shear thinning and suppression
of turbulent flow, which is related to flow induced changes in
chain conformation and orientation. These flow properties are of
great relevance in technical applications as thickeners, drag
reducers, and flow improvers, as well as in the production of
fiber-reinforced materials. A unique property of long chainlike
molecules is the formation of entanglement networks. Already at
low concentrations chain molecules start to overlap and entangle
to form a transient network. Chainlike molecules subjected to a
viscous shear gradient will orient in the flow, the instantaneous
angular velocity being a function of the orientation relative to
the local streamlines \cite{Orient:Shear}.
\newline
Meanwhile, researchers have been interested in the question of how
some material responds to an external mechanical load [2].
External loads cause liquids to flow, in Newtonian or various
types of non-Newtonian flows. Glassy materials, composed of
polymers, metals, or ceramics, can deform under mechanical loads,
and the nature of the response to loads often dictates the choice
of material in various industrial applications. In biological
systems, the response of proteins to external loads governs
aspects of cell adhesion and muscle function [3]. The nature of
all of these responses depends on both the temperature and loading
rate. As described by Eyring \cite{Eyr:JCP-4}, mechanical loading
lowers energy barriers, thus facilitating progress over the
barrier by random thermal fluctuations. The Eyring model
approximates the loading dependence of the barrier height as
linear. The Eyring model, with this linear barrier height
dependence on load, has been used over a large fraction of the
last century to describe the response of a wide range of systems
\cite{Rheo:Larson} and underlies modern approaches to biophysical
rupture processes \cite{Biop:Rup}, sheared glasses
\cite{Shear:Glass}, etc. To the best knowledge of the author, the
simplest model that makes a prediction for the rate and
temperature dependence of shear yielding is the rate-state Eyring
model of stress-biased thermal activation [3,7]. Structural
rearrangement is associated with a single energy barrier $E$ that
is lowered or raised linearly by an applied stress $\sigma$.
In glasses, the transition rates are negligible at zero stress.
Thus, at finite stress one needs to consider only the rate $R_{+}$
of transitions in the direction aided by stress.\newline The
linear dependence will always correctly describe small changes in
the barrier height, since it is simply the first term in the
Taylor expansion of the barrier height as a function of load. It
is thus appropriate when the barrier height changes only slightly
before the system escapes the local energy minimum. This situation
occurs at higher temperatures; for example, Newtonian flow is
obtained in the Eyring model in the limit where the system
experiences only small changes in the barrier height before
thermally escaping the energy minimum. As the temperature
decreases, larger changes in the barrier height occur before the
system escapes the energy minimum (giving rise to, for example,
non-Newtonian flow). In this regime, the linear dependence is not
necessarily appropriate, and can lead to inaccurate modeling. For
example, Li and Makarov \cite{Fold:Protein} have shown that there
is a nonlinear barrier height dependence in stretched proteins,
and that the assumption of a linear dependence in the analysis of
experimental results leads to inaccurate conclusions. To be
precise, at low shear rates ($\dot{\gamma} \le \dot{\gamma}_c$),
the system behaves as a power law shear-thinning material while,
at high shear rates, the stress varies affinely with the shear
rate. These two regimes correspond to two stable branches of
stationary states, for which data obtained by imposing either
$\sigma$ or $\dot{\gamma}$ exactly superpose. The transition from
the lower branch to the higher branch occurs through a stable
hysteretic loop in a stress-controlled experiment
\cite{Shear:Anneal}.
\newline
Note also that one prominent difference between the fluid motions
in nanodomain and those in macrodomain is the strong fluid-wall
interactions observed in nanoconduits. For example, as the
nanoconduit size decreases, the surface-to-volume ratio increases.
Therefore, various properties of the walls, such as surface
roughness, greatly affect the fluid motions in nanoconduits.
\newline
In this short paper,  we adopt the verified Eyring model [3-4] to
study the transport of shear-thinning fluids within corrugated
nanoannuli. To obtain the law of shear-thinning fluids for
explaining the too rapid annealing at the earliest time, because
the relaxation at the beginning was steeper than could be
explained by the bimolecular law, a hyperbolic sine law between
the shear (strain) rate : and (large) shear stress : $\tau$ was
proposed and  the close agreement with experimental data was
obtained. This model has sound physical foundation from the
thermal activation process [3-4] (Eyring [3] already considered a
kind of (quantum) tunneling which relates to the matter
rearranging by surmounting a potential energy barrier). With this
model we can associate the (shear-thinning) fluid with the
momentum transfer between neighboring atomic clusters on the
microscopic scale and reveals the atomic interaction in the
relaxation of flow with dissipation (the momentum transfer depends
on the activation shear volume, which is associated with the
center distance between atoms and is proportional to $k_B
T/\tau_0$ ($T$ is temperature in Kelvin, and $\tau_0$ a constant
with the dimension of stress). Thus, this model could be applied
to study transport of shear-thinning fluids in nanodomain [10].
\newline To consider the more realistic but complicated  boundary
conditions in the walls of nanoannulus, however, we will use the
boundary perturbation technique [11] to handle the presumed
wavy-roughness along the walls of nanoannuli. The relevant
boundary conditions along the wavy-rough surfaces will be
prescribed below.
\section{Physical Formulations}
We shall consider a steady transport of the (shear-thinning)
fluids in a wavy-rough nanoannulus of $r_2$ (mean-averaged outer
radius) with the outer wall being a fixed wavy-rough surface :
$r=r_2+\epsilon \sin(k \theta+\beta)$ and $r_1$ (mean-averaged
inner radius) with the inner wall being a fixed wavy-rough surface
: $r=r_1+\epsilon \sin(k \theta)$, where $\epsilon$ is the
amplitude of the (wavy) roughness, $\beta$ is the phase shift
between two walls, and the roughness wave number : $k=2\pi /L $.
Firstly, this fluid [3-4,10] can be expressed as
 $\dot{\gamma}=\dot{\gamma}_0  \sinh(\tau/\tau_0)$,
where $\dot{\gamma}$ is the shear rate, $\tau$ is the shear stress,
and $\dot{\gamma}_0$ is a function of temperature with the dimension
of the shear rate. In fact, the force balance gives the shear stress
at a radius $r$ as $\tau=-(r \,dp/dz)/2$. $dp/dz$ is the pressure
gradient along the flow (or tube-axis : $z$-axis) direction.\newline
Introducing
$\chi = -(r_2/2\tau_0) dp/dz$
then we have
 $\dot{\gamma}= \dot{\gamma}_0  \sinh ({\chi r}/{r_2})$.
As $\dot{\gamma}=- du/dr$ ($u$ is the velocity of the fluid flow
in the longitudinal ($z$-)direction of the nanoannulus), after
integration, we obtain
\begin{equation}
 u=u_s +\frac{\dot{\gamma}_0 r_2}{\chi} [\cosh \chi - \cosh (\frac{\chi r}{r_2})],
\end{equation}
here, $u_s$ is the velocity over the (inner or outer) surface of
the nanoannulus, which is determined by the boundary condition. We
noticed that Thompson and Troian [12] proposed a general boundary
condition for transport over a solid surface as
\begin{equation}
 \Delta u=L_s^0 \dot{\gamma}
 (1-\frac{\dot{\gamma}}{\dot{\gamma}_c})^{-1/2},
\end{equation}
where  $\Delta u$ is the velocity jump over the solid surface,
$L_s^0$ is a constant slip length, $\dot{\gamma}_c$ is the
critical shear rate at which the slip length diverges. The value
of $\dot{\gamma}_c$ is a function of the corrugation of
interfacial energy.  \newline With the boundary condition from
Thompson and Troian [12],  we can derive the velocity fields and
volume flow rates along the wavy-rough nanoannulus below using the
verified boundary perturbation technique [11]. The wavy boundaries
are prescribed as $r=r_1+\epsilon \sin(k\theta)$ and
$r=r_2+\epsilon \sin(k\theta+\beta)$ and the presumed steady
transport is along the $z$-direction (nanoannulus-axis direction).
\newline Along the outer boundary (the same treatment below could also be applied to
the inner boundary), we have
 $\dot{\gamma}=(d u)/(d n)|_{{\mbox{\small on surface}}}$.
Here, $n$ means the  normal. Let $u$ be expanded in $\epsilon$ :
 $u= u_0 +\epsilon u_1 + \epsilon^2 u_2 + \cdots$,
and on the boundary, we expand $u(r_0+\epsilon dr,
\theta(=\theta_0))$ into
\begin{displaymath}
u(r,\theta) |_{(r_0+\epsilon dr,\,\theta_0)} =u(r_0,\theta)+\epsilon
[dr \,u_r (r_0,\theta)]+ \epsilon^2 [\frac{dr^2}{2}
u_{rr}(r_0,\theta)]+\cdots=
\end{displaymath}
\begin{equation}
  \{u_{slip} +\frac{\dot{\gamma} r_2}{\chi} [\cosh \chi - \cosh (\frac{\chi
 r}{r_2})]\}|_{{\mbox{\small on surface}}}, \hspace*{6mm} r_0 \equiv
 r_1, r_2;
\end{equation}
where
\begin{equation}
 u_{slip}|_{{\mbox{\small on surface}}}=L_S^0 \{\dot{\gamma}
 [(1-\frac{\dot{\gamma}}{\dot{\gamma}_c})^{-1/2}]\}
 |_{{\mbox{\small on surface}}}, 
\end{equation}
Now, on the outer wall (cf. [11])
\begin{equation}   
 \dot{\gamma}=\frac{du}{dn}=\nabla u \frac{\nabla (r-r_2-\epsilon
\sin(k\theta+\beta))}{| \nabla (r-r_2-\epsilon \sin(k\theta+\beta))
|}.
\end{equation}
Considering $L_s^0 \sim r_1,r_2 \gg \epsilon$ case, we also presume
$\sinh\chi \ll \dot{\gamma}_c/\dot{\gamma_0}$.
With equations (1) and (5), using the definition of $\dot{\gamma}$,
we can derive the velocity field ($u$) up to the second order :
$u(r,\theta)$$=-(r_2 \dot{\gamma}_0/\chi) \{\cosh (\chi
r/r_2)$$-\cosh\chi\, [1+\epsilon^2 \chi^2 \sin^2 (k\theta+\beta)/(2
r_2^2)]+$$\epsilon \chi \sinh \chi \,
\sin(k\theta+\beta)/r_2\}$$+u_{slip}|_{r=r_2+\epsilon \sin
(k\theta+\beta)}$. The key point is to firstly obtain the slip
velocity along the boundaries or surfaces.
After lengthy mathematical manipulations, we obtain %
the velocity fields (up to the second order) and then we can
integrate them with respect to the cross-section to get the volume
flow rate ($Q$, also up to the second order here) :
%
 $ Q=\int_0^{\theta_p} \int^{r_2+\epsilon \sin(k\theta+\beta)}_{r_1+\epsilon \sin(k\theta)}
 u(r,\theta) r
 dr d\theta =Q_{slip} +\epsilon\,Q_{p_0}+\epsilon^2\,Q_{p_2}$.
In fact, the approximate (up to the second order) net volume flow
rate  reads $Q \equiv Q_{out} - Q_{in}$ which is the flow within
the outer (larger) wall : $Q_{out}$ without the contributions from
the flow within the inner (snaller) wall $Q_{in}$. \newline
\section{Results and Discussions} 
We shall demonstrate our results below. The wave number of
roughness is fixed to be $10$ (presumed to be the same for both
walls of the nanoannulus) for all figures here. Firstly, there is
an enhanced flow rate once the wavy-roughness is increasing [10].
This enhancement is rather significant especially when the forcing
(along the annulus-axis direction : $\chi/r_2$) is absent or zero
(purely slip flow). As the annulus size decreases, the
surface-to-volume ratio increases. Therefore, surface roughness
along both walls together with the slip-velocity boundary
condition, greatly affect the fluid motions between the corrugated
nanotubes. \newline Not that for an easy comparison, we select the
parameters to be  $r_2=2, r_1/r_2=0.5, L^0_s/r_1=1$;
$\dot{\gamma}_0/\dot{\gamma}_{c}=0.1$. For rather weakly
corrugations : $\epsilon=0.06 r_1$ the flow rate is monotonically
decreasing for $\chi/r_2 $ around 5. However, once we increase
$\chi/r_2$ to be around 7, the flow rate will be monotonically
increasing. $Q$ (the net volume flow rate) will firstly decrease
to a minimum and then keep increasing monotonically. This behavior
is almost the same for  $\beta$ (phase shift) being equal to
$\pi/4$, $\pi/2$ and $\pi$ (however, the effect of phase shift is
minor for very-small wavy-roughness [10]). It seems to us that,
however,  for very-small applied forcing there is a barrier
manifested by the shearing (of fluids along wavy-rough surfaces)
which leads to a minimum flow rate. As the forcing is large enough
and the barrier can be overcome then the transport (of fluids)
keeps increasing. The latter is something like a tunneling process
[3-4]!
\newline To examine the realistic effects of phase shift ($\beta$)
between the outer-wall wavy-roughness and the inner-wall one, we
fix the outer-wall and the inner-wall radii (to be 1 and 0.6 nm,
respectively). The small amplitude of wavy-roughness of both walls
is also fixed to be $0.06 r_1$. We consider two cases :
$\beta=\pi/4, \pi$ for the same $\dot{\gamma}_0=10000.0$
(s$^{-1}$) (cf. \cite{Shear:Rate}) with
$\dot{\gamma}_0/\dot{\gamma}_c =0.1$. The results are illustrated
in Fig. 2. We observe that for $\beta=\pi/4$ the net volume flow
rate $Q\equiv Q_{out}-Q_{in}$ is  decreasing as the forcing
increases (starting from zero but within a small range). The trend
for $\beta=\pi$, however, reverses! $Q$ for $\beta=\pi$ is
monotonically increasing and positive for larger forcing. Here,
the interesting observation is for small applied forcing ($4.5 \le
\chi/r_2 \le 7.5$, $r_2$ is fixed to be unity), once
$\beta=\pi/4$, the flow moves backward (along the annulus-axis
direction, even though the flow still moves forward once $\chi/r_2
\ge 8$ [14] as also evidenced in Fig. 1 since there is a minimum
flow rate for larger wavy-roughness for certain $\chi/r_2$ under
selected geometry). This result thus could be applied to the flow
control in nanofluidics.
\newline In brief summary, we have theoretically obtained the volume
flow rates (up to the second order) of Eyring-fluids inside the
wavy-rough annular nanotubes by using the verified fluid model
[3-4,10] and boundary perturbation method [11]. Our results show
that the wavy-roughness could tune the flow rate especially for
smaller forcing due to the larger surface-to-volume ratio and
slip-velocity effect. Meanwhile, the phase shift between the outer
and inner walls of nanoannuli could tune the transport of
shear-thinning fluids either forward or backward when the
wavy-roughness is larger enough as illustrated in Figure 2 here.
Our results could be applied to the flow control in nanofluidics
[15] as well as biofluidics [1].

\vspace*{25mm}

\psfig{file=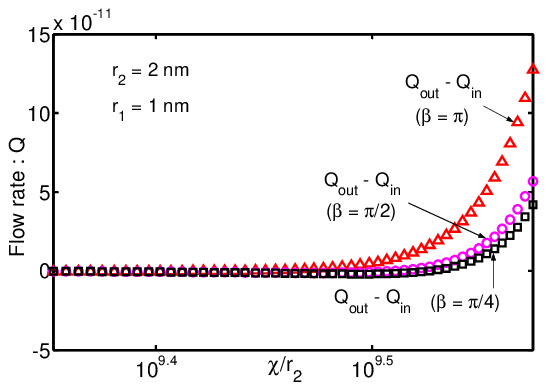,bbllx=-0.5cm,bblly=23cm,bburx=8cm,bbury=29cm,rheight=6cm,rwidth=6cm,clip=}

\begin{figure}[h]
\hspace*{10mm} Fig. 1 \hspace*{1mm} Calculated net volume flow
rates ($Q$) w.r.t. $\chi/r_2$ (forcing (along the $z$-axis
 \newline \hspace*{10mm} direction)
 per unit volume and referenced shear
stress). The mean outer radius is \newline \hspace*{10mm} $r_2=2$
(nm), and the mean inner one is $r_1 =1$ (nm).
 The slip length reads $L_s^0 = r_1$.\newline \hspace*{10mm}
$\epsilon$ (=0.06 $r_1$ here) is the
 amplitude of wavy roughness. The wave
number of \newline \hspace*{10mm} roughness ($k$) is $10$ here. We
demonstrate the effects of wavy-roughness via its phase shift
\newline \hspace*{10mm} $\beta$
(=$\pi/4,\pi/2, \pi$ here)  between the outer and inner walls of a
nanoannulus. For smaller $\beta$, \newline \hspace*{10mm} $Q$ is
smaller   w.r.t. the same $\chi/r_2$ (forcing). Meanwhile there is
a minimum flow rate
\newline \hspace*{10mm} (corresponds to a barrier). The unit of
volume flow rate is  $m^3/s$.
\end{figure} 

\newpage

\psfig{file=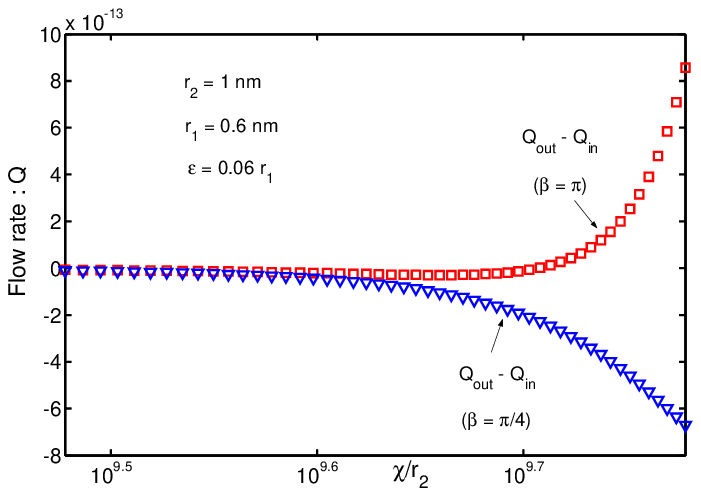,bbllx=-0.5cm,bblly=20.2cm,bburx=8cm,bbury=28.2cm,rheight=7.2cm,rwidth=7.2cm,clip=}

\begin{figure}[h]
\hspace*{10mm} Fig. 2 \hspace*{1mm} Calculated volume flow rates
($Q$) w.r.t. $\chi/r_2$ (forcing (along the $z$-axis
\newline \hspace*{10mm} direction)
 per unit volume and referenced shear
stress) for different phase shifts ($\beta$s). \newline
\hspace*{10mm} The mean outer radius $r_2=1$ nm, and the mean
inner radius $r_1=0.6$ nm. \newline \hspace*{10mm} The amplitude
of wavy-roughness $\epsilon$ (=0.06 $r_1$ here) is enlarged here
 and the wave \newline \hspace*{10mm}
number of roughness ($k$) is $10$ here. We consider two phase
shifts ($\beta=\pi/4, \pi$) and \newline \hspace*{10mm} the effect
is significant for the range of $\chi/r_2$ here. The unit of
volume flow rate is  $m^3/s$.
\newline \hspace*{10mm} $Q$ for $\beta=\pi/4$ is monotonically
decreasing (up to dimensionless $\chi/r_2\ge 8$ as $r_2$ is unity)
 \newline \hspace*{10mm} and there is backward transport for this
case. The trend for $\beta=\pi$ reverses. \newline \hspace*{10mm}
This result could be applied to the flow control in nanofluidics.
\end{figure}
%
\end{document}